\documentclass[letterpaper]{article} 
\usepackage{aaai25}  
\usepackage{times}  
\usepackage{helvet}  
\usepackage{courier}  
\usepackage{todonotes}
\usepackage[hyphens]{url}  
\usepackage{graphicx} 
\usepackage{natbib}  
\usepackage{caption} 
\usepackage{algorithm}
\usepackage{algorithmic}

\usepackage{amsmath}

\usepackage{newfloat}
\usepackage{listings}
\DeclareCaptionStyle{ruled}{labelfont=normalfont,labelsep=colon,strut=off} 
\floatstyle{ruled}
\newfloat{listing}{tb}{lst}{}
\floatname{listing}{Listing}
\title{GNN-Based Candidate Node Predictor for Influence Maximization in Temporal Graphs}

\author {
    Priyanka Gautam\textsuperscript{\rm *}, 
    Balasubramaniam Natarajan\textsuperscript{\rm *}, 
    Sai Munikoti\textsuperscript{\rm †}, \\
    S M Ferdous\textsuperscript{\rm †}, 
    Mahantesh Halappanavar\textsuperscript{\rm †}
}
\affiliations {
    \textsuperscript{\rm *}Electrical and Computer Engineering, Kansas State University, Manhattan, USA\\
    \textsuperscript{\rm †}Pacific Northwest National Laboratory, Richland, USA\\
    priyankagautam@ksu.edu, bala@ksu.edu, sai.munikoti@pnnl.gov, sm.ferdous@pnnl.gov, hala@pnnl.gov
}

\usepackage{bibentry}

\begin{document}

\maketitle

\begin{abstract}
In an age where information spreads rapidly across social media, effectively identifying influential nodes in dynamic networks is critical. Traditional influence maximization strategies often fail to keep up with rapidly evolving relationships and structures, leading to missed opportunities and inefficiencies. To address this, we propose a novel learning-based approach integrating Graph Neural Networks (GNNs) with Bidirectional Long Short-Term Memory (BiLSTM) models. This hybrid framework captures both structural and temporal dynamics, enabling accurate prediction of candidate nodes for seed set selection. The bidirectional nature of BiLSTM allows our model to analyze patterns from both past and future network states, ensuring adaptability to changes over time. By dynamically adapting to graph evolution at each time snapshot, our approach improves seed set calculation efficiency, achieving an average of 90\% accuracy in predicting potential seed nodes across diverse networks. This significantly reduces computational overhead by optimizing the number of nodes evaluated for seed selection. Our method is particularly effective in fields like viral marketing and social network analysis, where understanding temporal dynamics is crucial.

\end{abstract}

\section{Introduction}

In today’s constantly changing world, everything evolves rapidly, from traffic patterns to social networks. Roads that are clear in the morning may become congested by the afternoon, and the influential figures and posts on social media fluctuate constantly based on trends and shifting audience interests. These evolving dynamics are not random but follow underlying patterns that can be modeled as networks of interconnected nodes. It is critical to understand what is driving these changes, whether in traffic systems, social media, biology, or economics. A common thread in such challenges is the massive amount of time series data they generate. Understanding these data and key factors is crucial for predicting behavior, optimizing outcomes, and making informed decisions. One such fundamental problem in this context is influence maximization (InfMax), which is an important field of research in network analysis. The goal of InfMax is to identify a small subset of influential nodes, called seed nodes, that can maximize the spread of influence within a network. InfMax's applications are wide-ranging, from viral marketing to epidemic management and information dissemination.  However, solving InfMax is computationally challenging due to its NP-hard nature, making it computationally challenging to solve optimally \cite{kempe}. Early  InfMax approaches for static networks, proposed by Kempe et al. \cite{kempe}, and CELF \cite{celf}, tackled the challenge of finding the k-seed nodes to maximize influence with a given diffusion model, i.e., Independent Cascade or Linear Threshold Model, utilizing approximation techniques with Monte Carlo simulations. Subsequent research has focused on improving efficiency and scalability, incorporating additional constraints and objectives, and addressing real-world challenges through techniques such as CELF++, SCG, New Greedy-IC, UBLF, sketch-based algorithms \cite{celf++}, \cite{UBLF}, \cite{sketchbased}, and parallelization on shared and distributed memory CPUs \cite{Ripples, Preempt} and GPUs \cite{cuRipples}. 

Most of the existing InfMax algorithms are designed for static networks assuming a fixed network topology throughout the influence propagation process. These algorithms typically consider time-invariant influence probabilities between nodes, not accounting for temporal variations in relationships. 
In contrast, real-world networks are characterized by continuous evolution, with nodes and edges changing over time due to user interactions and network expansion.
The dynamic nature of networks necessitates adaptive algorithms that can cope with rapid changes in topology, connectivity, and node influence potential. Therefore, new frameworks have been proposed in the recent literature, including identifying network areas most affected by the changes \cite{topksteering}, modifying existing seed sets using replacement strategies \cite{UBI}, storing the recalculated shortest path \cite{Rossi2017MATIAE}, and employing snapshot-based prediction techniques \cite{snapshotprediction}. These innovations promise to transform influence maximization across various applications, including social network analysis and information dispersion. However, these algorithms pose some limitations: \cite{rossi_mati_2018} assumes new seed nodes will be in affected areas, overlooking potential influences from node deletions or emerging influential regions, while \cite{UBI}  makes strong assumptions about network properties and relies on propagation probabilities which may not be unknown at future instances. Additionally, approaches focusing on centrality measures \cite{centeralitymeasures-top-k} or structural properties \cite{infmax-global} often fail to capture the dynamic aspects of influence spread. Community-based algorithms \cite{localcommunity} become computationally expensive as graphs grow larger, exacerbating the NP-hard nature of the problem. This necessitates a critical need for a model that effectively captures both dynamic and structural properties of evolving networks while remaining computationally efficient.

To overcome these limitations, there has been a significant shift towards learning-based approaches in influence maximization research. Notable advancements include the snapshot prediction method for dynamic networks \cite{snapshotprediction}, which selects seed nodes based on both current and predicted future network states. \cite{Munikoti2024-yk} introduced GraMeR, a meta-reinforcement learning framework for generalized influence maximization using graph neural networks and its key components of formulating IM as a Markov decision process and using meta-learning for generalizability across different static graph types. \cite{PIANO} introduced PIANO, a deep reinforcement learning framework that demonstrates superior performance in efficiency and influence spread quality for various network snapshots that need retraining for each network. Taking inspiration from earlier learning-based developments, we offer a novel framework for optimizing dynamic influence. In this work, we combine  Graph Neural Networks (GNNs) to capture structural information with Long Short-Term Memory (BiLSTM) models to handle temporal dependencies in dynamic networks. This integrated approach aims to enhance the prediction and selection of influential nodes as network structures evolve. GNNs are popular for capturing structural information and spatial dependencies, while the BiLSTM model tackles temporal dependencies, making them ideal for predicting changes over time. Our framework takes input as a graph or network and predicts candidate nodes, i.e., potential seed nodes for each time snapshot, for maximum influence spread. The proposed approach is explained in Fig.\ref{fig: infmax-dynamic-flow}, and aims to provide an adaptive, efficient, and accurate influence maximization framework especially suited to dynamic network environments. The contributions of the paper are summarized below:
\begin{enumerate}
\item Propose a new dynamic framework that combines GNNs and LSTM models for real-time prediction of influential nodes in evolving networks.
\item Introduce a novel algorithm for candidate node prediction using influence capacity metrics and demonstrate the computational efficiency of this approach.
\item Demonstrate scalability across diverse network configurations using extensive simulations with 81\% - 98\% accuracy in predicting candidate node sets across various synthetic and real-world networks, which are subsequently used for seed set computation to maximize influence spread. 
\end{enumerate}



\section{Influence Maximization Problem and Dynamic Approach}

Infmax problem is centered around three essential components: the diffusion model, which dictates how influence propagates; optimization algorithms, which focus on selecting
the optimal set of seed nodes that maximize the influence
spread within the network. Finally, the network model itself
is crucial, as it provides the structural foundation on which influence spreads, whether in static or dynamic environments. As networks become increasingly dynamic, extending InfMax to temporal networks has become crucial to account for evolving network behaviors and structures.
\subsection{Diffusion Model}
The Independent Cascade (IC) and Linear Threshold (LT) models are extensively used and capture distinct diffusion mechanisms. In the linear model, nodes have activation thresholds that activate when the cumulative effect from neighbors reaches their threshold \cite{1978ThresholdMO}. While in the (IC) Model, activated nodes use probability to influence their neighbors in discrete time steps. Once activated, a node has a single opportunity to propagate influence in succeeding time steps \cite{kempe}. Recently, research has expanded beyond these models to develop advanced diffusion models that capture real-world complexities. These dynamic diffusion models address temporal changes, topic-specific influence, competitive processes, and continuous time spread. For instance, time-aware models allow influence probabilities to evolve, topic-aware models account for content-based spread patterns, and competitive models simulate concurrent diffusion processes. Continuous-time models also enable more realistic diffusion timing, and adaptive models adjust parameters based on network observations \cite{tracking_dynamic_survey_2018}.

\subsection{Optimization Algorithm} Without efficient optimization techniques, finding the optimal set of seed nodes would be computationally intractable for real-world networks, which often contain millions of nodes and edges. Given the NP-hard nature of influence maximization, greedy approximation algorithms are frequently used. These algorithms iteratively select seed nodes, often using Monte Carlo simulations to estimate influence spread. Simulation-Based Greedy Algorithms estimate influence by simulating multiple diffusion processes and selecting seed nodes based on the average spread \cite{kempe}\cite{celf}. Heuristic or proxy-based methods employ metrics like degree or centrality to approximate influence without extensive simulations, offering faster seed selection \cite{ifc_paper} \cite{degreheuristic} \cite{centeralitymeasures-top-k}. There are various heuristics-based metrics designed for
influence maximization, including degree, degree discount,
Katz centrality, etc. To facilitate effective seed selection, we apply the Influence Function Capacity (IFC), a heuristic combining local and global factors to evaluate node influence. Since our prior work \cite{Munikoti2024-yk} shows that the Influence function capacity (IFC) is an effective heuristic, we focus on IFC in the rest of the paper.
Lastly, Sketch-based approaches use efficient network summarization techniques to enable rapid influence estimation while maintaining approximation guarantees using precomputed paths \cite{sketchbased}.

\begin{figure*}

    \centering
    \includegraphics[width=\linewidth]{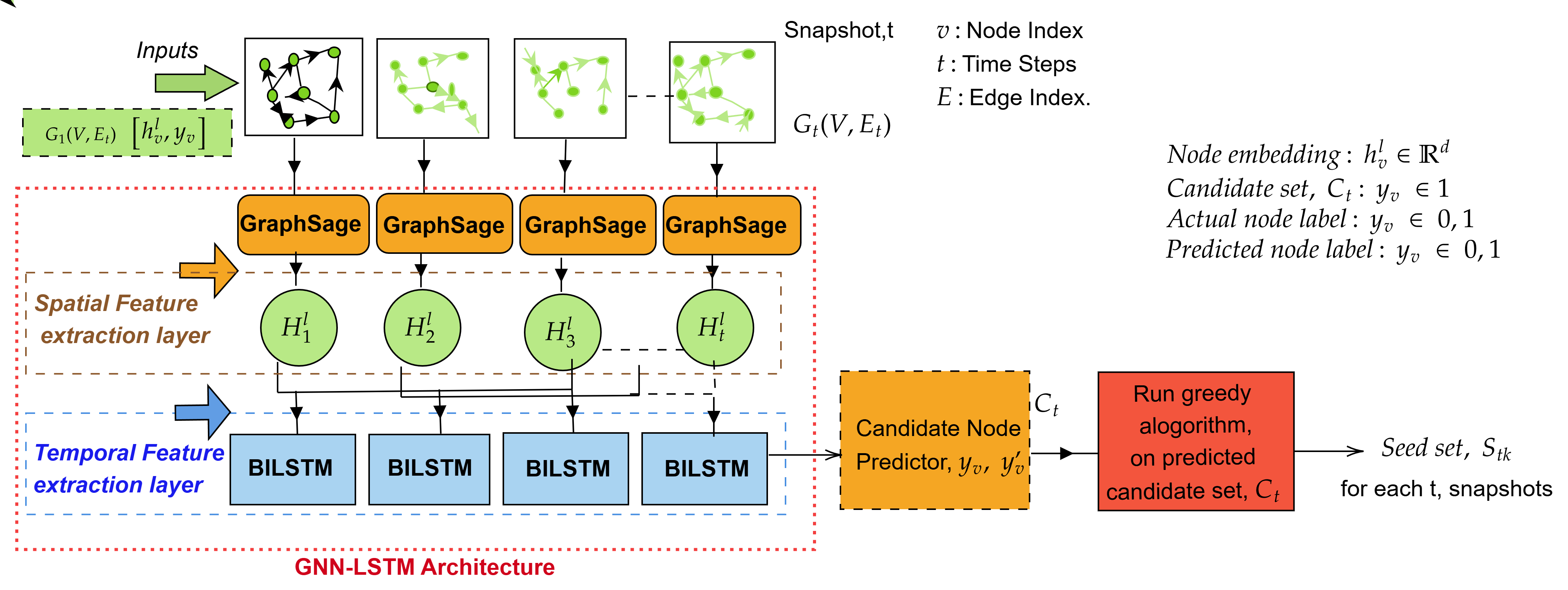}
    \caption{Overall flow diagram for the proposed framework to compute the seed set, i.e., most influential nodes. The framework considers graph snapshots for \( t \) time intervals that are processed through a GraphSAGE-based GNN to extract spatial features, generating embeddings \( H_t^l \) for each snapshot. These embeddings are then fed into a BI-LSTM, which uses a sequence-to-sequence model to capture temporal dependencies. For example in this setup, the BI-LSTM utilizes the embeddings from the previous d time intervals, i.e, d = 3 (\( H_{t-3}^l \), \( H_{t-2}^l \), \( H_{t-1}^l \)) to predict the embedding for the next interval (\( H_t^l \)). The candidate node predictor identifies a candidate set of influential nodes, and the Greedy algorithm selects the optimal seed set \( S_{tk} \) from this candidate set for each snapshot. 
    }
    \label {fig: infmax-dynamic-flow}
\end{figure*}

\subsection{Heuristic-Based Influence Capacity Scores}
 IFC  relies on local and global features of the graph to determine the influential seed nodes \cite{ifc_paper}. IFC calculates a node's influence potential by integrating its local connections and broader network position:
 \begin{itemize}
    \item \textbf{Local Influence \( I_L(x) \)}: 
 The local influence of a node \( x \) is bound within its two-hop neighborhood, as stated by \cite{Pei2014SearchingFS}. It can be computed as follows:
\begin{equation}
    I_L(x) = \sum_{j=2}^{j} I^j_L(x) = 1 + |N(x)|_G + \sum_{y \in N(x)} |N(y)|_{G \setminus \{x\}}
\end{equation}
where \( N(x) \) is the set of neighbors of node \( x \) in the network, \( |N(x)|_G \) represents the degree of node \( x \) in graph \( G \), and \( |N(y)|_{G \setminus \{x\}} \) is the degree of neighbor \( y \), excluding \( x \) itself. Alternatively, the local influence can also be calculated using influence weights \( w(x, y) \) for edges, as follows:
\begin{align}
    I_L(x) &= 1 + \sum_{y \in N(x)} w(x, y) \nonumber \\
           &\quad + \sum_{y \in N(x)} \sum_{z \in N(y)} w(x, y) \cdot w(y, z)
\end{align}
Here, \( w(x, y) \) is the influence weight of the edge between nodes \( x \) and \( y \), and \( N(y) \) is the set of neighbors of node \( y \) in the network. This formulation incorporates both direct connections and the weighted influence of two-hop neighbors to quantify the local influence of a node.
\end{itemize}
\begin{itemize}
    \item \textbf{Global Influence \( I_G(x) \)}: The global influence of a node \( x \) is estimated using its coreness score \( k_c(x) \), obtained via the \( k \)-shell decomposition method \cite{SEIDMAN1983269}. Coreness centrality, a global measure, identifies vital nodes by leveraging the network's global structure, given as follows:
\end{itemize}

\begin{equation}
I_G(x) = k_c(x) \times \left( 1 + \frac{D(x)}{D_N} \right) .
\label{eq:global_influence}
\end{equation}
Here, \(D(x) \) represents the degree of node \( x \) and \( D_N \) denotes the maximum degree in the network \( G \).
Combining these, the IFC score \(I(x)\) for a node \( x \) is given by:

\begin{equation}
I(x) = \frac{I_L(x)}{\max_{y \in V} I_L(y)} \times \frac{I_G(x)}{\max_{y \in V} I_G(y)} 
\label{eq:overall_influence_capacity}
\end{equation}

Utilizing this metric, we generate labels for each node to predict whether it is a candidate node or not. For example, nodes with an IFC score falling above a \(\alpha\) percentile are labeled as 1 (candidate node), while the rest are labeled as 0 (non-candidate node). The $\alpha$ values are chosen to balance capturing a significant portion of influential nodes while maintaining a manageable candidate set size. In summary, we leverage the IFC score to determine the candidate nodes which would eventually facilitate effective seed selections for the dynamic network.

\section{Proposed Methodology}

Our proposed framework (shown in Fig.\ref{fig: infmax-dynamic-flow}) integrates two key components to address influence maximization in dynamic network environments: (1) a heuristic-based approach for generating an initial candidate node set, and (2) a learning-based module that dynamically tracks and predicts candidate nodes by capturing both structural and temporal changes in the evolving graph. Unlike traditional methods that solely rely on heuristics for identifying candidate nodes, our objective is to predict these candidate nodes as the graph evolves. This predictive approach is critical, as computing heuristic scores such as Influence Capacity \cite{ifc_paper} for each node at each time snapshot is computationally expensive. Our approach aims to provide adaptive, efficient, and accurate influence maximization by leveraging the combined power of Graph Neural Networks (GNNs) \cite{graphSage} and Long Short-Term Memory (LSTM) \cite{Bilstm} models for real-time prediction. The following section details the candidate node predictor module and the GNN-LSTM architecture along with an implementation outline.

\subsection{Problem Formulation}

Let ${G}_t = (V_t, E_t)$ represent the graph at a specific time step \( t \), where \( V_t \) is the set of nodes and \( E_t \) is the set of edges. The dynamic network is modeled as a sequence of temporal graphs, \( \mathcal{G} = \{G_1, G_2, \ldots, G_T\} \), where \( T \) denotes the total number of time steps. At each variable time step \( t \), a candidate node predictor identifies a set of potential influencers, \( C_t \). From these candidates, a seed set \( S_t \), of size at most \( k \), is selected to maximize the expected influence, \( \sigma_t(S_t) \). As the network evolves over time, the influence maximization framework adapts to structural changes, including node additions, edge modifications, or deletions.

\subsection{Graph Neural Networks}

Graph Neural Networks (GNNs) are essential for capturing complex inter-dependencies in dynamic networks through message-passing and aggregation. In each GNN layer, nodes iteratively aggregate features from their neighbors to update embeddings, 
\begin{align}
    \mathbf{h}_N^{(l)}(v) &\leftarrow \text{AGG} \left( \mathbf{h}_u^{(l-1)}, \forall u \in N(v) \right)
\end{align}
Here, \( \mathbf{h}_u^{(l)} \) represents the feature vector of node \( u \) at layer \( l \), encoding its own features and aggregated information from its neighbors, where \(\mathcal{N}_S(v)\) represents sampled neighbors of node \(v\) and \( N(v) \) denotes the neighbors of node \( v \). In this work, we used GraphSAGE, which employs a sampling-based aggregation mechanism, that uses two important steps message passing and aggregation. The update rule for node \( v \) at layer \( l \) can be formulated as:


\begin{equation}
\begin{split}
    \mathbf{h}_v^{(l)} &\leftarrow \sigma \Big( \mathbf{W}^{(l)} \cdot \text{CONCAT} \Big( \mathbf{h}_v^{(l-1)}, \\
    &\quad \text{AGG} \left( \{ \mathbf{h}_u^{(l-1)} \mid \forall u \in N(v) \} \right) \Big) \Big)
\end{split}
\end{equation}

where, \(\mathbf{h}_v^{(l)} \) is the hidden representation of node \( v \) at layer \( l \) and  \( \mathbf{W}^{(l)} \) is the learnable weight matrix for layer \( l \) and \( \sigma \) is a non-linear activation function (e.g., ReLU). At each layer concatenation, of the current node’s representation \( \mathbf{h}_v^{(l-1)} \) with the aggregated messages from its neighbors  is done to generate the final embedding. The graph convolution operation in the matrix form can be expressed as:

\begin{equation}
\mathbf{H}^{(l+1)} = \sigma\left(\tilde{\mathbf{D}}^{-\frac{1}{2}} \tilde{\mathbf{A}} \tilde{\mathbf{D}}^{-\frac{1}{2}} \mathbf{H}^{(l)} \mathbf{W}^{(l)}\right),
\end{equation}

where \(\mathbf{H}^{(l)}\) represents the node representations at layer \(l\), and \(\mathbf{H}^{(l+1)}\) denotes the updated representations for the next layer. \(\tilde{\mathbf{A}}\) is the adjacency matrix with added self-loops to account for the node's own features, and \(\tilde{\mathbf{D}}\) is its corresponding degree matrix for normalization. The learnable weight matrix \(\mathbf{W}^{(l)}\) facilitates feature transformation, while the activation function \(\sigma(\cdot)\) introduces non-linearity. These techniques empower the model to adapt to dynamic states and refine candidate selection by leveraging graph embeddings and IFC scores.

The candidate node predictor uses the input graphs to construct the dynamic graph representation, as  $f(G_t)$ incorporating both temporal and structural information. Each node's initial feature vector \(\mathbf{h}_v^0\) is augmented with a five-dimensional vector \([d_v, \bar{d}_v, 1, 1, 1]\), where \(d_v\) is the node degree and \(\bar{d}_v\) is the average degree of its neighbors:
\begin{equation}
\bar{d}_v = \frac{1}{|N(v)|} \sum_{u \in N(v)} d_u
\end{equation}
This augmentation enhances the representation of local and neighborhood-level properties. The augmented vectors are processed through GNN layers to generate embeddings \(\mathbf{h}_v^l\), encapsulating both structural and temporal dynamics. These embeddings are used to classify nodes as candidates for influence maximization, reducing the need for repeated heuristic computations and enabling efficient identification of influential nodes as the network evolves.


\begin{table*}[t]
    \centering
    \begin{tabular}{l|r|r|r|r}
        \hline
        \textbf{Dataset} & \textbf{Nodes} & \textbf{Edges (across all snapshots)} & \textbf{Graphs} & \textbf{Accuracy (\%)} \\
        \hline
         email-Eu-core-temporal-Dept1 & 320 & 61,046 & 19 & 92.50 \\
        \hline
        Random-Barabasi-Albert & 1,000 & 10,696 & 20 & 97.86 \\
        \hline
        College-Msg & 1,900 & 12,821 & 20 & 87.66 \\
        \hline
        Random graph-ER & 2,500 & 128,599 & 20 & 81.36 \\
        \hline
        Autonomous Systems-Oregon & 11,492 & 22,724 & 9 & 96.62 \\
        \hline
    \end{tabular}
    \caption{Performance of the GNN-based Candidate Node Predictor across
    datasets with varying graph properties.}
    \label{tab:gnn_node_predictor}
\end{table*}


\subsection{Candidate Node Predictor: GNN-LSTM Model}

The GNN-LSTM model forms the core of the candidate node predictor, enabling the integration of spatial and temporal features to dynamically predict potential seed nodes for influence maximization. As illustrated in Fig.\ref{fig: infmax-dynamic-flow},  the model operates in two primary phases: spatial feature extraction and temporal feature modeling.

\subsubsection{Spatial Feature Extraction:}

In the first phase, graph snapshots for \( t \) time intervals are processed through a GraphSAGE-based GNN \cite{graphSage}. The input to this component includes the graph structure \( G_t \) and the initial feature vectors \( h_v^0 \) for each node \( v \) at time \( t \). The GraphSAGE model aggregates information from a node’s local neighborhood up to \( l\) layers generating embeddings \( h_v^l \) for each node. This embedding captures the structural properties of each node by considering its immediate \( l\) hops neighbors, providing a rich representation of the node's local graph structure. The embeddings \( h_v^l \) for each node in a graph snapshot are then aggregated to form the overall graph embedding \( \mathbf{H}_t^l \), which encapsulates the structural features of the entire graph at time \( t \). This graph embedding serves as a compact representation of the spatial features necessary for influence maximization.


\subsubsection{Temporal Feature Modeling}
In the second phase, the graph embeddings \( \mathbf{H}_t^l \) generated by the GNN are fed into a Long Short-Term Memory (LSTM) network to capture temporal dependencies across multiple time snapshots. To improve temporal modeling, we employ a Bidirectional LSTM (BiLSTM) \cite{Bilstm}, which processes the embeddings in both forward and backward directions, providing a more comprehensive understanding of the graph's dynamic evolution. Specifically, the BiLSTM utilizes a sequence-to-sequence model where the embeddings from the previous \(d\) time intervals are used to predict the embeddings for the next interval \( \mathbf{H}_{(d+1)}^l \). It enables the model to anticipate changes in node influence as the network dynamics unfold. The GNN-LSTM model finally predicts i.e., candidate node set \( C_t \) for each graph. 

By considering both the structural properties of the graph and its temporal evolution, this prediction process enhances the framework’s ability to adapt to changes in the network, ensuring that the selected seed nodes are optimized for influence maximization across evolving conditions. Fig.\ref{fig: infmax-dynamic-flow}, illustrates the GNN-LSTM model architecture, where the spatial features are extracted by the GraphSAGE model and then processed through the BiLSTM to predict candidate nodes set for influence maximization, the candidate node predictor results is summarized in Table.\ref {tab:gnn_node_predictor}. These candidate nodes set is used to find the influential nodes for seed set computations. The overall process of identifying seed nodes for the dynamic graphs is summarized in Algorithm. \ref{alg:dynamic_infmax}.

\begin{algorithm}[tb]
\caption{Dynamic Influence Maximization using Candidate Node Predictor and Greedy Algorithm}
\label{alg:dynamic_infmax}
\begin{algorithmic}[1]
\STATE \textbf{Input:} Dynamic network snapshots $\mathcal{G} = \{G_1, G_2, \ldots, G_T\}$, Seed set size $k$; Influence Function $\sigma(S)$; Candidate set $\{C_t\}(C_t =  f(G_t) )$
\STATE \textbf{Output:} Seed sets $\{S_{tk}\}$ for each time step $t$

\FOR{each time step $t = 1$ to $T$}
    \STATE Extract graph $G_t = (V_t, E_t)$ from $\mathcal{G}$
    \STATE Initialize $S_{tk} \leftarrow \emptyset$
    
    \FOR{$i = 1$ to $k$}
        \STATE $u^* \leftarrow \arg \max_{u \in C_t \setminus S_{tk}} \left( \sigma(S_{tk} \cup \{u\}) - \sigma(S_{tk}) \right)$
        \STATE $S_{tk} \leftarrow S_{tk} \cup \{u^*\}$
    \ENDFOR

\ENDFOR

\RETURN Seed sets $\{S_{tk}\}$
\end{algorithmic}
\end{algorithm}

\section{Experiments \& Results}
\subsection{Datasets: Real-world and Synthetic Graphs}
In this study, we use a combination of real-world and synthetic datasets to evaluate our proposed model. The real-world datasets were sourced from the SNAP dataset collection \cite{snapnets}. Specifically, we used the email-Eu-core-temporal-Dept1 dataset i.e email network created from European research institutions. It contains 320 nodes, and $61,046$ edges, over a periods of 803 days converted into 19 snapshots based on months. We also included the College-MSg dataset with $1,900$ nodes, and $12,821$ edges, which creates 20 snapshots. In addition to these real-world datasets, we generated synthetic datasets using the Erdős-Rényi (ER) and Barabasi-Albert (BA) generative models to simulate various network structures and dynamics with both edge addition and deletions. Furthermore, we used Random graph-ER dataset: with $2,500$ nodes, and $128,599$ edges spread across 20 snapshots. These datasets provided a diverse set of network configurations to thoroughly evaluate the performance and robustness of our approach.

To generate dynamic network snapshots, we utilized datasets with distinct temporal properties. The Email-Eu-Dept network naturally exhibits dynamic behavior, as communication between individuals at one time step (\( t_1 \)) may not persist at a later time step (\( t_2 \)), resulting in the disappearance of edges over time. Similarly, the Autonomous Systems (AS) Oregon dataset captures the evolution of AS peering relationships inferred from Oregon route views, reflecting the dynamic nature of internet routing topology. Graph snapshots were created based on temporal resolutions, such as months, to capture gradual changes in network structure. For synthetic graphs, edge addition probabilities ranged between 0.1 and 0.5 to model network growth, while edge deletion probabilities ranged between 0.5 and 0.7 to represent structural decay or significant drops in connectivity. These probabilistic modifications ensured that the generated snapshots reflected both growth and decline patterns commonly observed in dynamic networks.

To facilitate reproducibility and promote further research, the code and data supporting this study are available at \url{https://github.com/Priyankagautam08/DynamicGraphInfluenceMaximization-AAAI2025}{GitHub}.


\subsection{Model Training and Experimental Setup}

We trained our model on real-world dynamic social network datasets with the following hyperparameters: learning rate of 0.001, 0.01, batch size of 4, the sequence length of BILSTM, $d=3$, and we utilize the value of $\alpha$ = 40 \% i.e we choose the modes falling in the 60th percentile and higher to be labeled as candidate node set. The hidden size determines the dimensionality of the hidden state vector of size = 128, 256, and finally trained the model from ranging to 200-400 epochs based on different datasets. The experiments were conducted on a system equipped with a 12th Gen Intel(R) Core(TM) i9-12900K 3.20 GHz processor and an NVIDIA GeForce RTX 3090 GPU. The model's performance was evaluated using an accuracy score. Our approach was compared against the traditional Greedy algorithm, highlighting improvements in terms of computational efficiency.

\subsection{GNN Candidate Node Predictor Results}
The candidate node predictor is trained across various classes of datasets with different graphs and node sizes ranging from 1000 to 11k nodes and types (real-world vs. synthetic datasets). The ground truth or labels of node classes are obtained using the influence capacity metric. 
We assessed the accuracy of our GNN-based candidate node predictor across various datasets, highlighting its ability to effectively capture dynamic network changes. Table \ref{tab:gnn_node_predictor} summarizes the accuracy results, demonstrating strong performance across most datasets, with particularly high accuracy on the Random-Barabasi-Albert and Autonomous systems datasets.

\begin{figure}[!ht]
    \centering
    \includegraphics[width=0.48\textwidth]{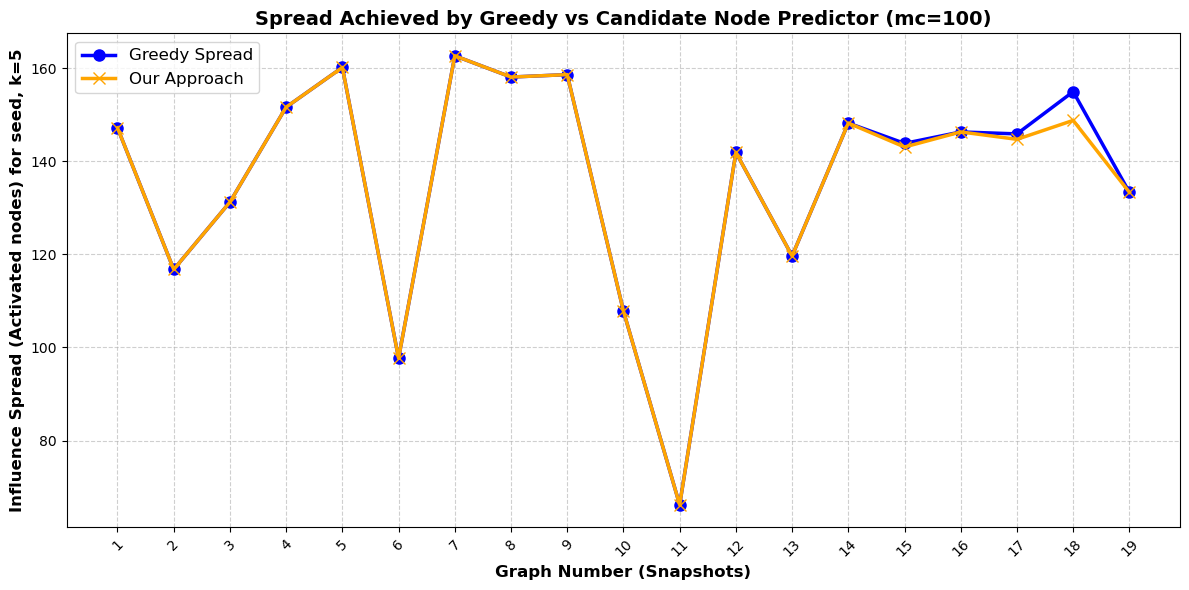}
    \caption{Comparison of influence spread achieved by the Greedy algorithm versus our candidate node predictor approach across 19 graph snapshots on Email core temporal department messages network. The spread is measured using Monte Carlo simulations with 100 iterations (\( mc = 100 \)) and a seed set size \( k = 5 \). The dips in influence spread are due to changes in the network structure (as few old edges are dropping and new edges are forming.)}
\label{fig:spread_comparison}
\end{figure}

\subsection{Comparative Analysis of Influence Spread and Computational Efficiency}
\begin{figure}[!ht]
    \centering
    \includegraphics[width=0.48\textwidth]{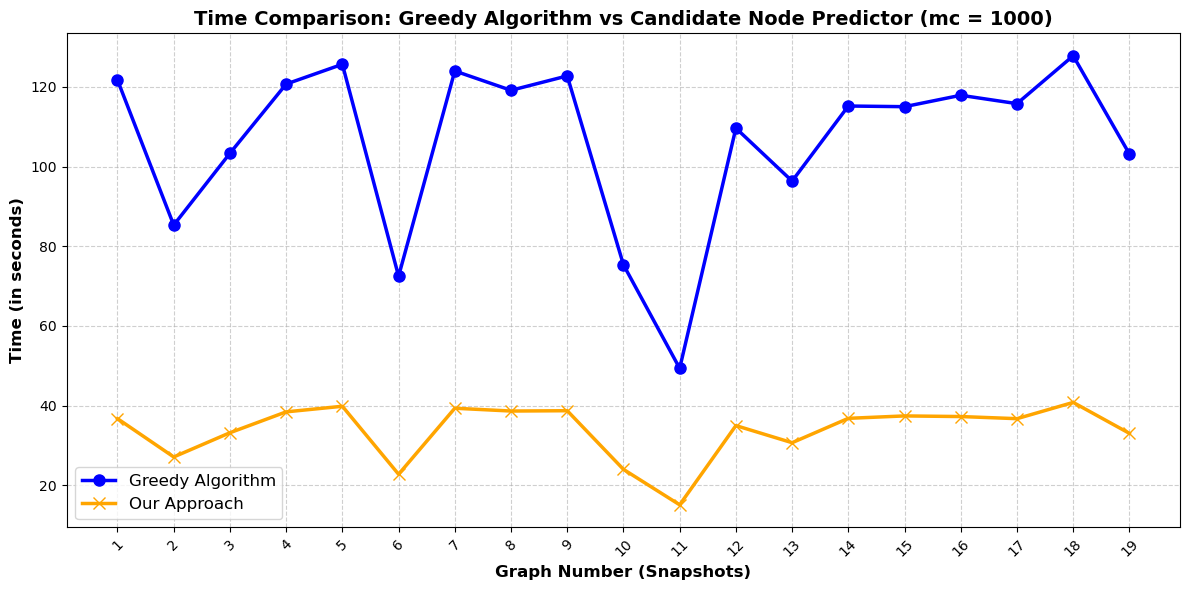}
    \caption{Comparison of computational time taken by the Greedy algorithm versus our candidate node predictor approach across 19 graph snapshots on Email core temporal department messages network, measured in seconds. The experiments were conducted using Monte Carlo simulations with 100 iterations (\( mc = 100 \)). 
    }
\label{fig:time_comparison}
\end{figure}

We compared the performance of our proposed candidate node selection method against the traditional Greedy algorithm focusing on both the spread of influence and computational efficiency. As depicted in Figure~\ref{fig:spread_comparison}, our approach consistently matches the spread achieved by the Greedy algorithm. The results indicate that our approach closely matches the performance of the Greedy algorithm across most snapshots, demonstrating its effectiveness in maintaining high influence spread while potentially reducing computational overhead. Furthermore, Figure~\ref{fig:time_comparison} illustrates a significant reduction in computational time across all graphs. Our method is nearly two times faster in computation time relative to the Greedy algorithm, highlighting its efficiency. These results underscore the effectiveness of our approach in dynamic environments where rapid response and adaptability are crucial.

\subsection{Analysis of Dips in Spread and Compute Time}

The observed dips in influence spread and computational time around snapshots 9 to 11 (as shown in Figures \ref{fig:spread_comparison} and \ref{fig:time_comparison}) can be attributed to structural changes in the directed graph. During these snapshots, while the number of nodes remained constant at 320, the number of edges reduced notably, from 820 to 773, and in snapshots 6 to 7 number of edges increased from 808 to 927. Concurrently, the number of strongly connected components decreased from 101 in Snapshot 11 to 96 in Snapshot 13. There is a reduction in connected components as well which suggests that previously fragmented subgraphs are merging into larger, more cohesive structures, while some connections are lost and new ones are formed. In a directed graph, this consolidation can significantly reduce overall complexity, leading to more efficient influence propagation and computation. According to our experiments and conclusions, the decreases in processing time represent the algorithms' ability to handle fewer isolated regions, whilst the brief decrease in impact spread indicates the network's shift to a more unified structure. Furthermore, the observed dips are caused by a loss of node influence, network saturation, and declining returns as the graph grows or get denser and, in certain cases, sparser. Interestingly, in a few graphs despite the observed dips, the centrality of key nodes remained stable or increased, indicating that the continued decrease in influence spread is not due to a loss of node influence. Instead, this decrease can be attributed to network saturation, where the influence quickly reaches most nodes, and diminishing returns on spread as the network becomes more interconnected, reducing the impact of additional seed nodes.

\begin{figure}[!ht]
    \centering
    \includegraphics[width=0.48\textwidth]{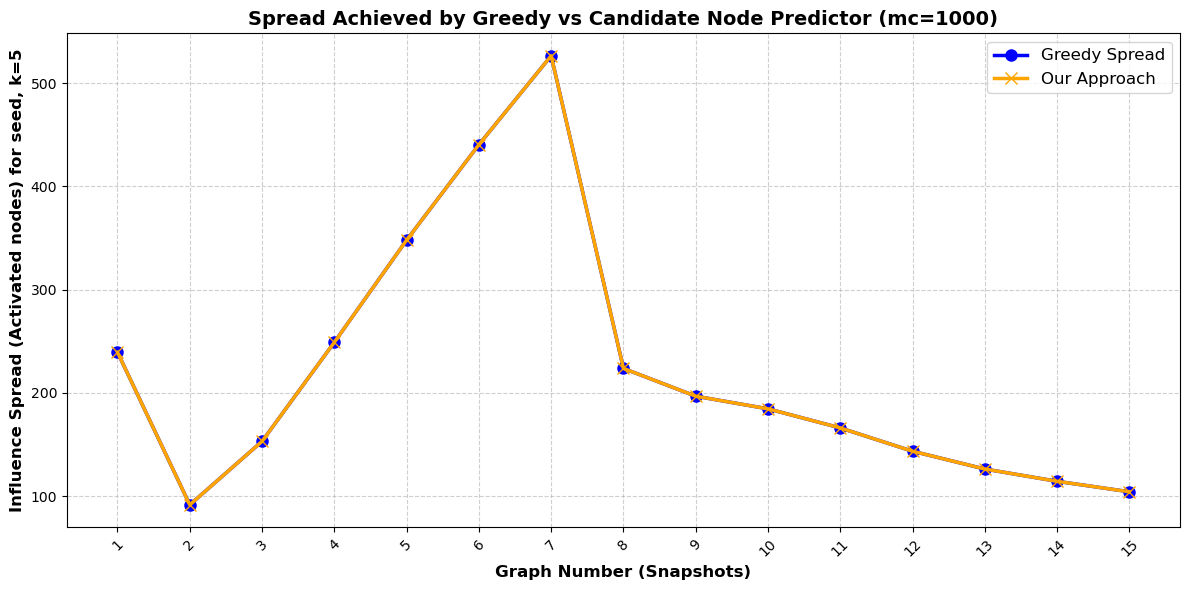}
    \caption{Comparison of influence spread achieved by the Greedy algorithm versus our candidate node predictor approach across 15 graph snapshots on Synthetic Random-Barabasi-Albert network with 1000 nodes. The spread is measured using Monte Carlo simulations with 1000 iterations (\( mc = 1000 \)) and a seed set size \( k = 5 \).}
\label{fig:spread_comparison2}
\end{figure}

\begin{figure}[!ht]
    \centering
    \includegraphics[width=0.48\textwidth]{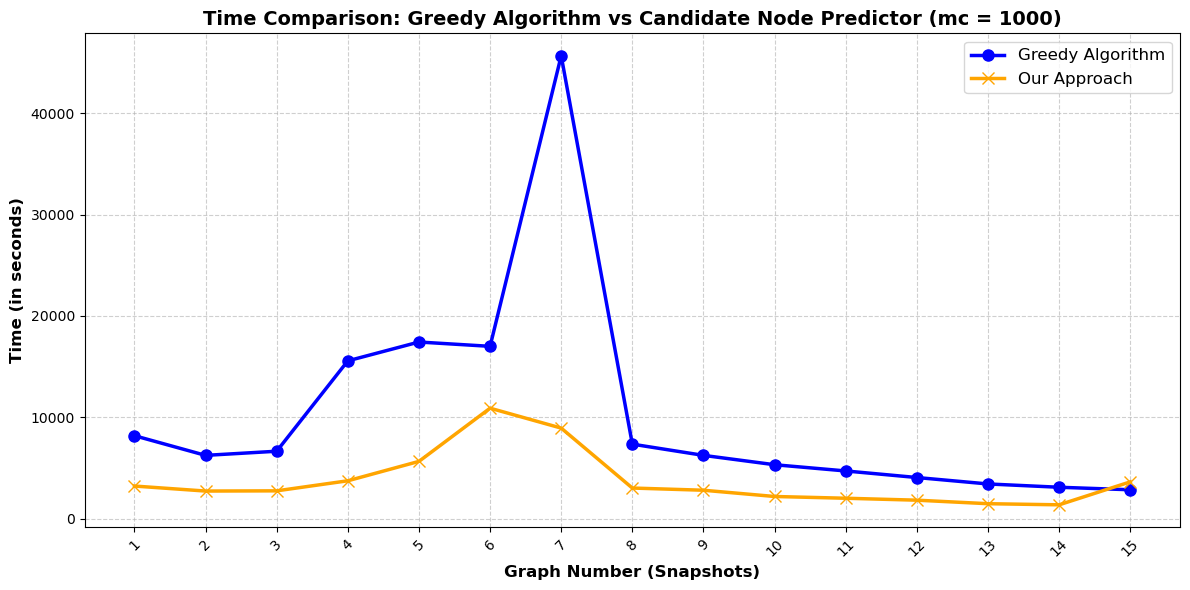}
    \caption{Comparison of computational time taken by the Greedy algorithm versus our candidate node predictor approach across  15 graph snapshots on Synthetic Random-Barabasi-Albert network with 1000 nodes. The spread is measured using Monte Carlo simulations with 1000 iterations (\( mc = 1000 \)) and a seed set size \( k = 5 \). 
    }
\label{fig:time_comparison2}
\end{figure}

\subsection{Performance on Synthetic and Real-world Inputs}
We evaluated our candidate node selection method on both synthetic and real-world networks. For real-world data, we focused on the Email core temporal department messages network, analyzing 19 graphs to capture temporal and structural dynamics. Results are shown in Figures \ref{fig:spread_comparison} and \ref{fig:time_comparison}. For synthetic networks, we used the Random Barabasi-Albert model with 1000 nodes and 13032 edges across 15 instances, as seen in Figures \ref{fig:spread_comparison2} and \ref{fig:time_comparison2}. Our method demonstrated strong performance, with key observations: (1) Spread Equivalence with the greedy algorithm, (2) Adaptability in predicting outcomes despite spread drops, and (3) Robustness to network changes, ensuring its suitability for real-world applications.


\subsection{Performance Across Scales}

To further evaluate the efficiency and effectiveness of our GNN-based candidate node predictor, we tested it on networks of varying sizes. This included both real-world networks and synthetic networks such as BA and Erdős-Rényi (ER) models with sizes ranging from 200 to 2500 nodes. These varied size tests ensured that our approach scales well and maintains performance across small to large network scenarios. Our GNN candidate node predictor demonstrated varying but generally strong performance across different network sizes. As shown in Table \ref{tab:gnn_node_predictor}, the model achieved remarkable accuracy ( above 90\%) on large networks with greater than 500 nodes. For medium-sized networks of 1000-2500 nodes, it maintained a solid performance with above 80\% accuracy. Interestingly, the model showed remarkable resilience to scaling, achieving 96.62\% accuracy on the large Autonomous Systems - Oregon network with 11,492 nodes. This suggests that our approach can effectively handle small and large-scale network structures.

\section{Summary \& Conclusions}
This paper presents a novel approach for dynamic influence maximization, combining a heuristic-based learning method with network analysis techniques. The proposed methodology leverages the strengths of Graph Neural Networks (GNNs) and Bi-directional Long Short-Term Memory (BiLSTM) models to predict influential nodes in real-time. A heuristic-based influence scoring mechanism, coupled with a dynamic graph representation, enables our framework to adapt effectively to changes in the network over time. 
Our candidate node prediction algorithm achieve significant improvements in terms of both accuracy and computational efficiency when selecting the optimal seed set. The effectiveness of our approach has been validated through extensive simulations, demonstrating over 81\% to 98\% accuracy in influence spread and strong scalability across diverse network structures. Additionally, the flexibility of our framework allows for integration with various diffusion models, and heuristic approaches making it broadly applicable to real-world scenarios. 

In the future, we plan to decouple the detection and prediction of influence nodes to develop explicit strategies dedicated to detecting changes in dynamic network structures, allowing us to better understand the underlying shifts. Concurrently, we will design adaptive influence maximization strategies that can be adjusted based on the insights gained from the change detection process. This separation will enable us to refine each aspect independently, ultimately improving the framework's responsiveness to network evolution and optimizing its performance in dynamic environments. Another key direction is to emphasize persistent influence by tracking node stability over time and employing dynamic optimization techniques to maximize long-term influence. These enhancements extend the framework’s scalability and effectiveness across diverse real-world networks.


\section*{Acknowledgment}

This work is supported by the National Science Foundation under Award No. OIA-2148878 with matching support from the State of Kansas through the Kansas Board of Regents, and by the U.S. Department of Energy through the Exascale Computing Project (17-SC-20-SC) (ExaGraph) at the Pacific Northwest National Laboratory.

\bibliography{aaai25}

\end{document}